\documentclass[twocolumn,tighten,twocolappendix]{aastex631}
\shorttitle{What is Missing from the Local Stellar Halo?}
\shortauthors{Sharpe et al.}
\graphicspath{{./}{figures/}}

\usepackage{xspace} 
\usepackage{natbib} 
\usepackage[capitalize]{cleveref}
\usepackage{rotating}
\usepackage{threeparttable}
\usepackage{float}
\usepackage{graphicx}

\defcitealias{Bullock+Johnston2005}{BJ05}

\newcommand{\package}[1]{\tt{#1}}

\newcommand{\Msun}{\ensuremath{\,\mathrm{M_\odot}}\xspace} 
\newcommand{\Gyr}{\ensuremath{\,\mathrm{Gyr}}\xspace} 

\begin{document}

\title{What is Missing from the Local Stellar Halo?}

\author[0000-0001-8225-8969]{Katherine Sharpe}
\affiliation{Center for Astrophysics | Harvard \& Smithsonian, 60 Garden Street, Cambridge, MA 02138, USA}
\email{ksharpe@college.harvard.edu, rnaidu@mit.edu}

\author[0000-0003-3997-5705]{Rohan P. Naidu}
\thanks{NASA Hubble Fellow}
\affiliation{Center for Astrophysics | Harvard \& Smithsonian, 60 Garden Street, Cambridge, MA 02138, USA}
\affiliation{MIT Kavli Institute for Astrophysics and Space Research, 77 Massachusetts Avenue, Cambridge, MA 02139, USA}
\author[0000-0002-1590-8551]{Charlie Conroy}
\affiliation{Center for Astrophysics | Harvard \& Smithsonian, 60 Garden Street, Cambridge, MA 02138, USA}


\begin{abstract}

The Milky Way's stellar halo, which extends to $>100$ kpc, encodes the evolutionary history of our Galaxy. However, most studies of the halo to date have been limited to within a few kpc of the Sun. Here, we characterize differences between this local halo and the stellar halo in its entirety. We construct a composite stellar halo model by combining observationally motivated N-body simulations of the Milky Way's nine most massive disrupted dwarf galaxies that account for almost all of the mass in the halo. We find that (1) the representation by mass of different dwarf galaxies in the local halo compared to the whole halo can be significantly overestimated (e.g., the Helmi Streams) or underestimated (e.g., Cetus) and (2) properties of the overall halo (e.g., net rotation) inferred via orbit integration of local halo stars are significantly biased, because e.g., highly retrograde debris from Gaia-Sausage-Enceladus is missing from the local halo. Therefore, extrapolations from the local to the global halo should be treated with caution. From analysis of a sample of 11 MW-like simulated halos, we identify a population of recently accreted ($\lesssim5$ Gyrs) and disrupted galaxies on high angular momenta orbits that are entirely missing from local samples, and awaiting discovery in the outer halo. Our results motivate the need for surveys of halo stars extending to the Galaxy's virial radius.
\end{abstract}


\section{Introduction} \label{sec:introduction}

In $\Lambda \rm CDM$ cosmology, galaxies grow hierarchically, with smaller systems continuously merging into more massive galaxies \citep[e.g.,][]{White+Frenk1991}. Our clearest view into this hierarchical assimilation comes from the stellar halo of the Milky Way, which is almost entirely comprised of debris from accretion events \citep[e.g.,][]{dimatteo19, Mackereth20, Naidu+2020}. Families of halo stars that arrived as part of the same galaxy retain similar phase space properties (e.g., energies, angular momenta, actions; \citealt{Brown2005,Gomez2013,Simpson2019}) as well as shared chemical abundance patterns \citep[e.g.,][]{Lee+2015, Cunningham+2022}. With detailed chemodynamical data, it is challenging \citep[e.g.,][]{Jean-Baptiste+2017J} but possible to determine which populations of stars were originally associated with the same dwarf galaxy merger event. From satellite galaxies being currently disrupted \citep[e.g., Sagittarius;][]{Ibata+1994} to stellar populations fully integrated into the halo \citep[e.g., Thamnos;][]{Koppelman+2019}, dwarf galaxies in all stages of accretion can be found in and around the stellar halo, encoding our Galaxy's assembly history.

The stellar halo is difficult to study directly due to its small relative mass ($\approx1\%$ of the Galaxy's stellar mass; \citealt{Deason+2019}), large spatial extent, and the necessity to collect full 6D phase-space and abundance data to reconstruct its history with high fidelity. Thanks to numerous spectroscopic surveys (e.g., RAVE, \citealt{RAVE}; SEGUE, \citealt{SEGUE}; LAMOST, \citealt{LAMOST}; GALAH, \citealt{GALAH}; APOGEE, \citealt{APOGEE}; H3, \citealt{Conroy+2019}; and the \textit{Gaia} mission, \citealt{Gaia-Collaboration+2018}), we have detailed chemodynamical parameters for thousands of stars in the stellar halo. However, owing to observational feasibility, the vast majority of halo stars studied in depth are located in the solar neighborhood, within a few kiloparsecs of the Sun (the ``local halo").

From the local halo, it is possible to infer the properties of the more distant halo. An important technique is integrating local halo stars' orbits and analyzing their properties based on their orbital apocenters, which is the maximum distance they reach from the Galactic center. This method is, for example, applied to determine the net rotation of the outer halo, which may provide evidence as to its method of formation \citep[e.g.,][]{Carollo+2007,Schonrich+2011,Beers+2012,Helmi+2017}.

However, we know that the local halo is unrepresentative of the stellar halo in its entirety. For instance, stars associated with the prominent dwarf galaxies Cetus \citep[e.g.,][]{Newberg+2009} and Sagittarius are not found within the solar neighborhood; thus, any extrapolations from local samples cannot capture any information about these two accreted dwarf galaxies. Apocenter analyses do not accurately capture the more radially extreme stars in an accreted dwarf. Accreted stars are most likely to be found close to their apocenters \citep[e.g.,][]{Deason+2018}, meaning stars on large-apocenter orbits are least likely to be found near the Sun. Additionally, there is a degree of selection bias when determining which stars within the solar neighborhood actually belong to the stellar halo and which are associated with the Galactic disk (e.g., stars that are retrograde with respect to the disk are more likely to be classified as halo stars). 

In this paper, we aim to precisely determine what differences might exist between extrapolations from the local halo and the stellar halo as a whole. We begin by constructing a composite model of the Milky Way's stellar halo, described in Section\,\ref{sec:method} and Section\,\ref{sec:MWCOMPOSITE}. Then, in Section\,\ref{sec:local_halo_results}, we study the relative representation of stars in the local halo compared to the stellar halo in its entirety for all accreted dwarfs in the composite model. We also perform apocenter analysis, after integrating the orbits of local halo stars, to compare apocenter properties to those of whole halo stars, focusing on net prograde/retrograde motion. In Section\,\ref{sec:MW_context}, we compare the composite halo with the eleven halo simulations from \cite{Bullock+Johnston2005, Robertson+2005, Font+2006} (hereafter \citetalias{Bullock+Johnston2005}). Finally, in Section\,\ref{sec:conclusion} we summarize our conclusions and present a few final remarks.


\begin{deluxetable*}{lrrrrr}[t!]
\label{tab:summary}
\tabletypesize{\footnotesize}
\tablecaption{Summary of Accreted Dwarf Galaxies in the Milky Way Stellar Halo Composite Model.}
\tablehead{
\colhead{Dwarf Galaxy} & \colhead{$\log(M_{\star}/M_{\rm{\odot}})$} & 
\colhead{$N_{\rm{\star}}$} & \colhead{Mass per Particle (\Msun)} & \colhead{Time Unbound (\Gyr)} &
\colhead{Simulation Source}}
\startdata
\vspace{-0.2cm}  \\
Sagittarius & 8.5 (8.8) & 200,000 & 3,155 & 0 & \cite{Vasiliev+2021}\\
GSE & 8.7 & 50,060 & 9,988 & 9 & \cite{Naidu+2021}\\
Kraken & 8.3 & 3,002 & 63,291 & 12 & \citetalias{Bullock+Johnston2005}: Halo 3, ID 91\\
Helmi Streams & 8.0 & 100,000 & 1000 & 5-8 (6.5) & \cite{Koppelman+2019}\\
Sequoia & 7.2 & 12,512 & 1,267 & 7 & \citetalias{Bullock+Johnston2005}: Halo 1, ID 51\\
Wukong/LMS-1 & 7.1 & 15,580 & 808 & 8 & \cite{Malhan+2021}\\
Cetus & 7.0 & 200,000 & 50 & 5 & \cite{Chang+2020}\\
Thamnos & 6.7 & 10,256 & 489 & 11 &  \citetalias{Bullock+Johnston2005}: Halo 6, ID 66\\
I'itoi & 6.3 & 2,645 & 754 & 10 & \citetalias{Bullock+Johnston2005}: Halo 4, ID 75\\
\enddata
\tablecomments{Objects are listed in order of decreasing total stellar mass. Stellar masses are sourced from \citet{Naidu+2022} for all objects except for Kraken, for which we use \citet{Kruijssen20}. The total stellar mass of Sagittarius is shown in the parenthetical; we cut approximately $3\times10^8\Msun$ about the center of the Sagittarius dwarf galaxy as it is often considered separately from analysis of the stellar halo. $N_*$ is the number of particles from each simulation. Due to differences in simulation resolution for these nine objects, we weight our analysis on the mass per particle (the total stellar mass of each dwarf galaxy divided by the number of particles in its N-body simulation).  We include the ``time unbound", representing the disruption epoch for each dwarf galaxy when it ceased to be a gravitationally bound object. For the stated range on the Helmi Stream's time unbound, we adopt the median of the range reported in \cite{Koppelman+2019}. For Kraken, Sequoia, Thamnos, and I'itoi, simulations are selected from \citetalias{Bullock+Johnston2005} based on similarities in $L_z$, total energy, and 3D Galactocentric distance.}

\end{deluxetable*}

\section{Method} \label{sec:method}

\subsection{Simulation Sources}
\label{sec:sim_sources}

In this section, we describe how our composite Milky Way stellar halo is constructed, as well as the set of simulations from \citetalias{Bullock+Johnston2005} we use to place the composite model in context. 

We assemble a composite Milky Way stellar halo from observationally motivated simulations representing nine accreted dwarf galaxies: Gaia-Sausage-Enceladus \citep[GSE, e.g.,][]{Belokurov+2018, Haywood+2018, Helmi+2018,Bonaca+2020, Feuillet+2021, Naidu+2021, Buder+2022}, Sagittarius \citep[e.g.,][]{Ibata+1994, LawMajewsk2010, Johnson+2020, Vasiliev+2021}, Kraken (\citealt{Kruijssen+2019, Kruijssen20, Massari+2019, Forbes+2020, Horta2021,Pfeffer+2021,Naidu+2022rproc}; though note that the stars and clusters associated with Kraken may belong to the proto-Milky Way recently characterized in \citealt{Belokurov22,Conroy22,Myeong22,Rix+2022}), the Helmi Streams \citep[e.g.,][]{Helmi+1999, Koppelman+2019, Limberg+2021,Matsuno2022}, Sequoia \citep[e.g.,][]{Myeong+2019, Matsuno+2019, Matsuno+2021, Monty+2020, Aguado+2021}, Wukong/LMS-1 \citep[hereafter Wukong, e.g.,][]{Naidu+2020, Yuan+2020, Malhan+2021, Malhan+2022, Shank+2022}, Cetus \citep[e.g.,][]{Newberg+2009, Chang+2020, Thomas+2021, Yuan+2021}, Thamnos \citep[e.g.,][]{Koppelman+2019, Lovdal+2022, Ruiz-Lara+2022}, and I\'itoi \citep[e.g.,][]{Naidu+2020, Naidu+2022}. These nine systems together comprise almost all known $M_{\star}\gtrsim 10^{6} M_{\rm{\odot}}$ accreted systems, and nearly all of the total stellar halo ($\approx10^{9}M_{\rm{\odot}}$, \citealt{Deason+2019}) by mass (see Table 1 of \citealt{Naidu+2022} for stellar mass estimates of individual dwarfs). Models of these systems therefore provide a realistic approximation of the Milky Way for the issues we are interested in. 

We note that several lower mass ($<10^{6} M_{\rm{\odot}}$) dwarf galaxies \citep[e.g.,][]{Shipp+2018, Ji+20, Bonaca+2021, Tenachi+2022, Dodd+2022, Chandra+2022} are predicted \citep[e.g.,][]{Robertson+2005,Deason+2016,Fattahi+2019} and observed \citep[e.g.,][]{Naidu+2020,Helmi20,An21} to contribute a small minority of the stellar halo. While we do not include such systems in our composite Milky Way model, they are captured in the \citetalias{Bullock+Johnston2005} simulated halos we analyze, and the general trends we report apply to them.

We use star particles from five tailor-made N-body simulations, representing GSE \citep[][]{Naidu+2021}, Sagittarius \citep{Vasiliev+2021}, the Helmi Streams \citep{Koppelman+2019}, Wukong/LMS-1 \citep[hereafter Wukong,][]{Malhan+2021}, and Cetus \citep[][]{Yuan+2021}. We note that for Sagittarius, we cut out the intact core and keep the remaining $\approx 3 \times 10^8 \Msun$ found within the disrupted tails. However, Sagittarius is shown in its entirety as a part of Figures \ref{fig:mw_halo_xy} and \ref{fig:mw_halo_xz}.

Four of the dwarfs we consider --- Kraken, Sequoia, Thamnos, and I'itoi --- have not yet had dedicated simulations run. We select representative models for these four objects from the simulated dwarf galaxies in the halos from \citetalias{Bullock+Johnston2005} on the basis of similarities in phase and real space. We begin by selecting the sample of dwarfs from \citetalias{Bullock+Johnston2005} with similar total mass, average z-component angular momentum ($L_z$), and total energy to each of the four dwarfs. We then visually compare the $L_z-E_\mathrm{tot}$ to observational data. Where multiple simulations had comparable phase space distributions, similarities in average galactic radius were also considered. 

Later in the paper (Section \ref{sec:MW_context}) we also compare the eleven \citetalias{Bullock+Johnston2005} simulated halos in their entirety to our composite Milky Way stellar halo. These halos follow cosmologically motivated accretion histories for Milky Way mass galaxies. Each halo has on the order of 100 accreted dwarf galaxies. 

The key properties of the individual simulations comprising the composite Milky Way halo are summarized in Table \ref{tab:summary}. To account for the differences in resolution between simulations, we weight each particle by distributing the total stellar mass of each accreted dwarf galaxy evenly across all of the simulation's N particles (see ``Mass per Particle" in Table \ref{tab:summary}).

\subsection{Dynamical Properties and Orbit Integration}
\label{sec:orbit_integration}

We integrate orbits for our composite stellar halo particles in the default \texttt{gala} Milky Way gravitational potential \citep{Price-Whelan+2018}. This potential consists of a spherical nucleus and bulge, a disk, and a spherical dark matter halo \citep{Bovy2015}. For the additional eleven halo simulations, halo parameters reported in \citetalias{Bullock+Johnston2005} are implemented via \texttt{gala}'s composite potential functionality. Due to the differences between our adopted fiducial Milky Way potential and the potentials for the \citetalias{Bullock+Johnston2005} simulations, the four simulated dwarfs selected to represent Thamnos, Sequoia, I'itoi, and Kraken are rescaled to retain their approximate positions in phase space --- all particle velocities are scaled by the ratio of the circular velocities at their average radius in the Milky Way potential to that of their \citetalias{Bullock+Johnston2005} potential ($v_\mathrm{circ, MW}$/$v_\mathrm{circ, BJ05}$).
  
We use the default Galactocentric frame from \texttt{Astropy} v4.2.1. This frame is right-handed, and the origin is placed at the galactic center. Prograde motion is represented by a negative $z$-component of angular momentum ($L_{z}<0$). Orbit integrations are performed on stars within the solar neighborhood via \texttt{gala}. The orbits of all stars in the solar neighborhood are integrated forward by 5\Gyr. This replicates how the outer halo is often explored in the literature via local halo samples \citep[e.g.,][]{Carollo+2007}.

\section{Results} \label{sec:results}

\subsection{Overview of the Composite Milky Way Stellar Halo} \label{sec:MWCOMPOSITE}

\begin{figure*}[h!]
\centering
\includegraphics[width=1\textwidth]{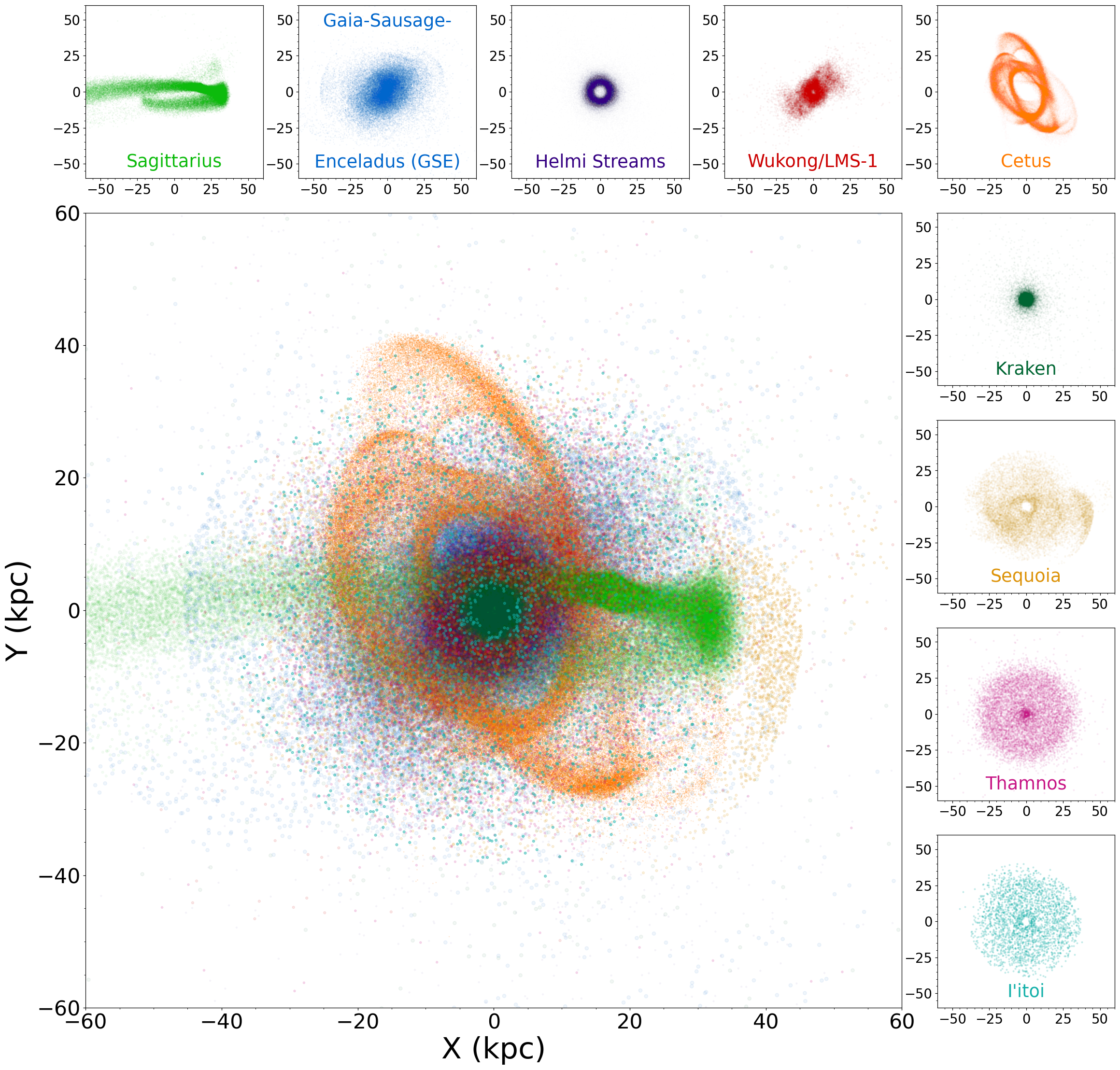}
\caption{Composite Milky Way stellar halo, displayed in the $XY$ plane. Each accreted dwarf galaxy is shown in color as its own panel and in the large, composite panel. The four panels directly to the right of the large composite plot are those selected from \citetalias{Bullock+Johnston2005}. In the large panel, point size and opacity are roughly scaled to the power 3/4 and 1/3 of mass per particle, respectively, with slight scaling variations for visual clarity.}
\label{fig:mw_halo_xy}

\end{figure*}

\begin{figure*}[ht!]
\centering
\includegraphics[width=1\textwidth]{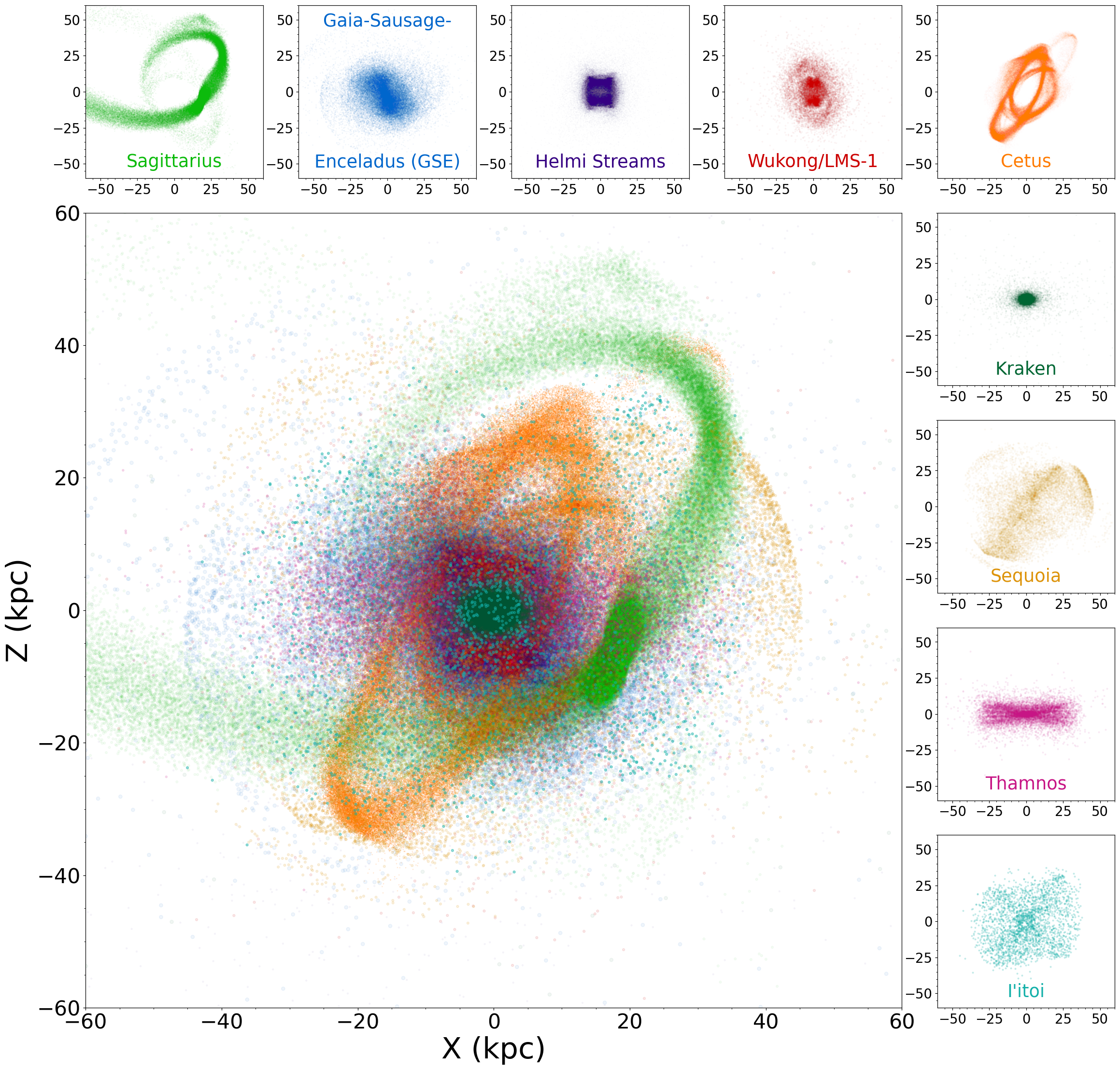}
\caption{Composite Milky Way stellar halo, displayed in the $XZ$ plane. See caption of Figure \ref{fig:mw_halo_xy}.}
\label{fig:mw_halo_xz}
\end{figure*}

The analysis in this paper centers around the composite simulated Milky Way stellar halo described in Section \ref{sec:sim_sources}. We display this composite halo as a projection onto the $XY$ and $XZ$ planes in Figures \ref{fig:mw_halo_xy} and \ref{fig:mw_halo_xz}, respectively. Each of the simulated dwarfs are shown in color in a mini-panel surrounding the composite view. Simulations selected from \citetalias{Bullock+Johnston2005} are located along the right. 

We observe three types of distributions for these simulated dwarfs. In order of increasing relaxation: (1) stream-like, including Sagittarius and Cetus; (2) spheroidal, including GSE, the Helmi Streams, Sequoia, Wukong, and I'itoi; and (3) compact spheroidal or flattened, including Kraken and Thamnos. Based on this figure, it appears that the solar neighborhood is most likely to represent stars from spheroidal-type dwarfs, due to their more homogeneous distribution. Likewise, stream-like dwarfs and the most compact of the spheroidal dwarfs are most likely to be absent from the solar neighborhood, since their physical extent does not intersect the position of the Sun.

We quantitatively explore the relative representation of accreted dwarfs in the local halo in Section \ref{sec:local_halo_results}, and study their angular momenta distributions in Section \ref{sec:MW_context}.

\subsection{Examining Bias in the Local Halo}
\label{sec:local_halo_results}

\begin{figure*}[t!]
\centering
\includegraphics[width=0.95\linewidth]{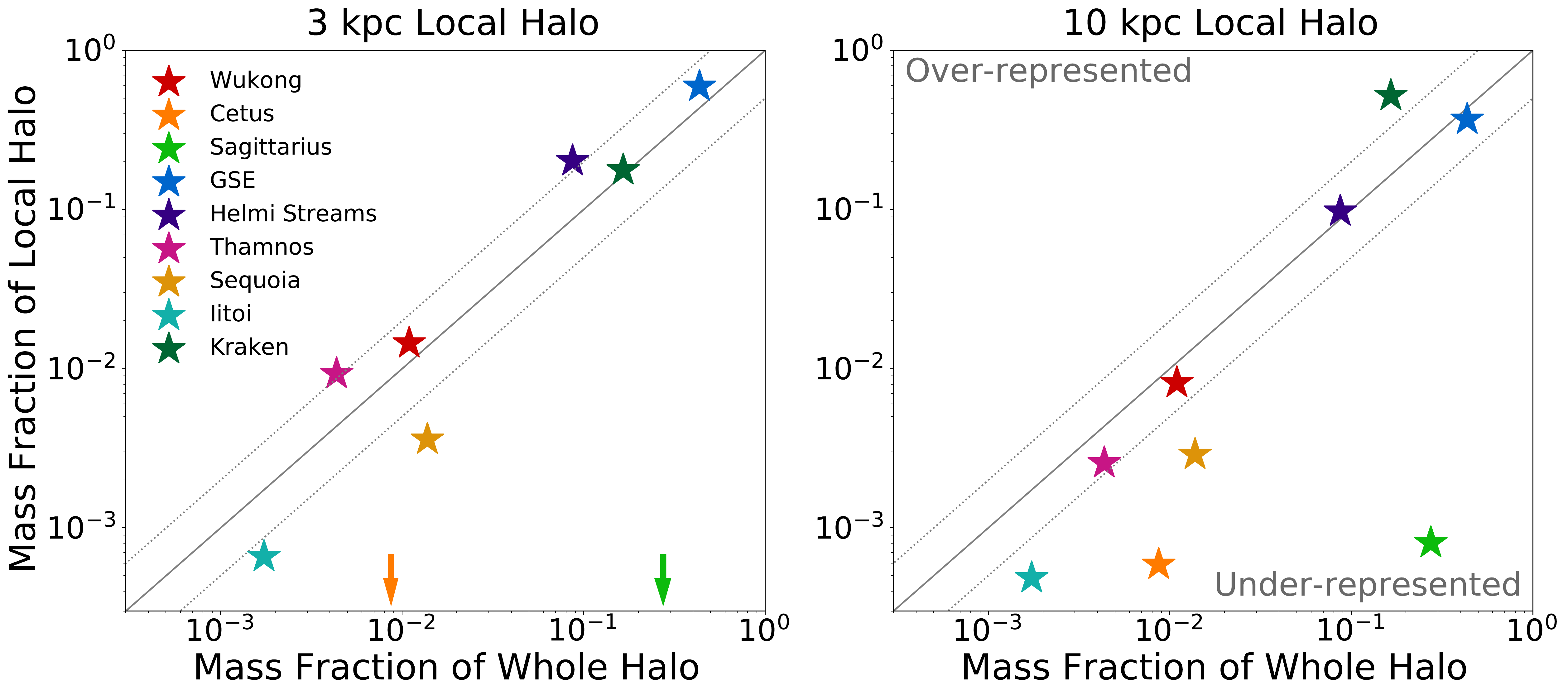}
\caption{Comparison of the fractional composition of the whole halo to that of the local halo for each object. The left panel shows this comparison for a local halo with radius 3\,kpc, while the right shows the same for an extended local halo with radius 10\,kpc. The central gray line is where all points would fall if representations were equal between the local halo and the whole halo. The upper and lower gray dashed lines show a local halo representation $2\times$ and $0.5\times$ the representation of the whole halo, respectively. Note that the two entirely unrepresented objects with a local halo fraction of zero, Cetus and Sagittarius, are shown as upper limits, and that the above plots are in log scale.}
\label{fig:frac_halo_sn}
\end{figure*}

We begin our analysis by considering to what degree each accreted dwarf galaxy is represented within the local halo sample. In an unbiased sample, we expect the relative representation of each accreted dwarf by mass to approximately equal its whole halo mass fraction. 

In Figure \ref{fig:frac_halo_sn}, we compare the mass fractions of each dwarf galaxy within the local halo to those of the whole halo. This comparison is made for local halo with radii 3\,kpc (left), and 10\,kpc (right). Equal representation between the local halo and the whole halo falls along the central gray 1:1 line. Cetus and Sagittarius are absent in the 3\,kpc local halo, and are shown as arrows pointing towards zero. 

Representation within the local halo does not approximate the mass fraction of each of the composite halo dwarf galaxy sources. For example, both Cetus and Sagittarius are completely unrepresented in the 3\,kpc local halo, and are still under-represented in the 10\,kpc local halo. Additionally, I'itoi and Sequoia are under-represented by more than a factor of two. Meanwhile, Kraken is over-represented in the 3\,kpc local halo by more than a factor of two --- this over-representation is increased in the 10\,kpc local halo as the solar neighborhood then includes the galactic center. Note that in practice, it can be extremely challenging to probe the halo around the Galactic center due to extinction. This situation is changing rapidly thanks to \textit{Gaia} -- for instance, \citet{Rix+2022} recently used the low-resolution XP spectra from DR3 to reveal a metal poor population which is very compact around the Galactic center. 

Now, we examine the properties of local halo stars at their apocenters. In the literature, apocenter analysis is used to extrapolate observations of the local halo to the distant halo --- local halo stars are taken to be distant stars on an interior part of their orbit. However, there are some challenges associated with this type of analysis. Even in a homogeneous stellar halo, it is more difficult to capture stars with large apocenters in a local halo sample --- these stars spend much of their orbits at large radii, and thus are unlikely to be found near the Sun. 

\begin{figure}[t!]
\centering
\includegraphics[width=0.95\linewidth]{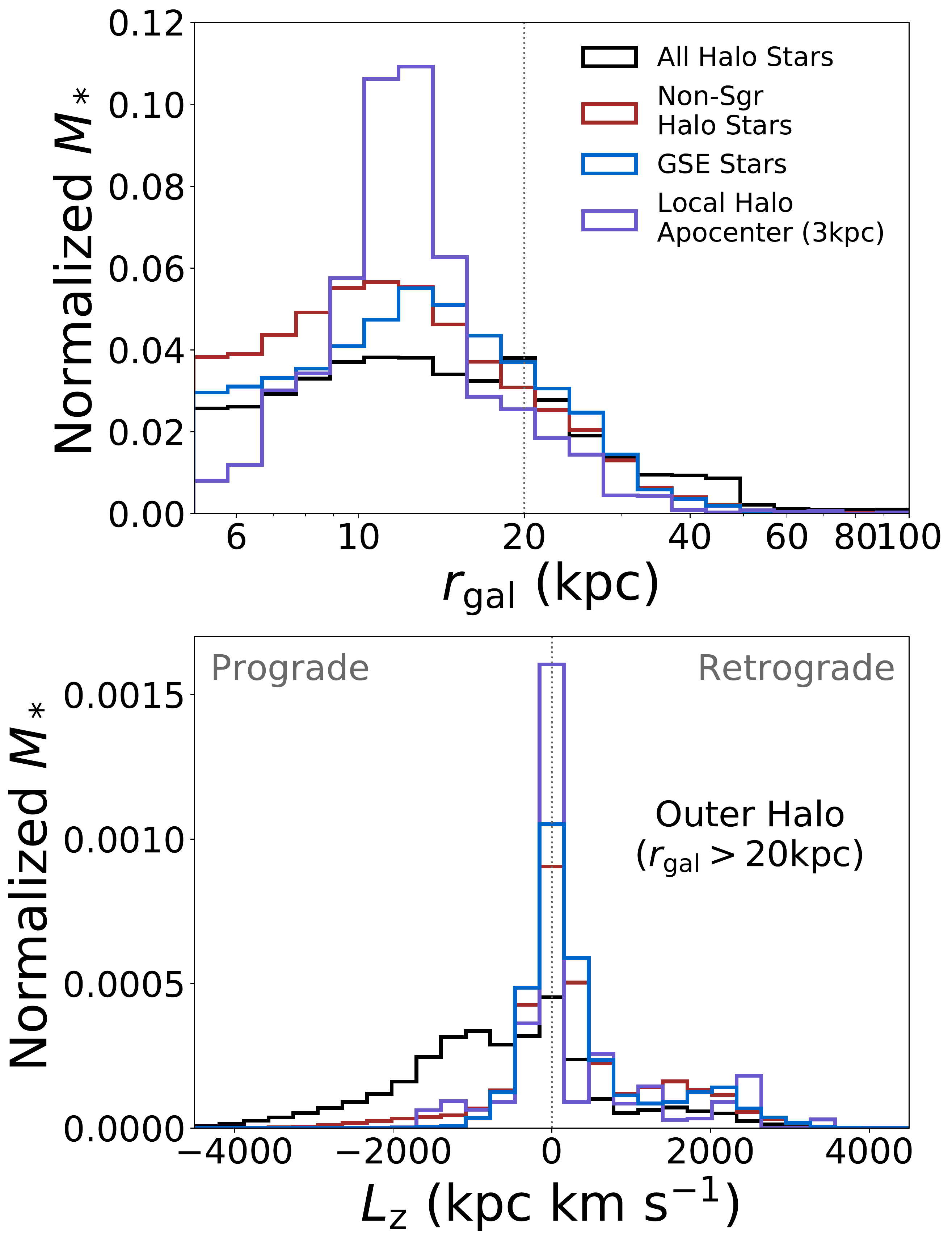}
\caption{Histograms comparing the properties of the present day whole halo (in black), whole halo without Sagittarius (brown), and GSE (blue) to those of the local halo's apocenters (for a 3\,kpc solar neighborhood in purple). \textbf{Top:} The relative stellar mass at each Galactocentric radius. The gray dashed line at 20\,kpc denotes the boundary beyond which we classify stars as being in the `outer halo'. At radii greater than about 20\,kpc, we see a smaller fraction of stars from local halo apocenters when compared to the entire population of halo stars. The local halo's apocenter distribution peaks at $\approx15-20$ kpc, corresponding to GSE's inner orbital apocenter \citep[e.g.][]{Naidu+2021, Han+2022}. \textbf{Bottom:} $L_z$ for local halo stars with apocenters in the outer halo (beyond 20\,kpc) to those of the actual halo beyond 20\,kpc, using the same color scheme as the upper panel. The local halo is significantly more radial, i.e., concentrated at $L_z\approx0$ than the actual halo.}
\label{fig:r_lz_hist}
\end{figure}

In the upper panel of Figure \ref{fig:r_lz_hist}, we show the relative amount of stellar mass at each radius approximated based on apocenters of local halo stars. We also show GSE, the stellar halo without Sagittarius, and the stellar halo in its entirety. We show non-Sagittarius halo stars since Sagittarius is the most massive component beyond about 30\,kpc, and it is useful to see to what degree the results depend on this one dwarf. Similarly, GSE dominates the local halo, and it is interesting to isolate its effect.

The most striking feature in the upper panel of Figure \ref{fig:r_lz_hist} is that local halo stars are selectively probing $10-15$\,kpc. This peak corresponds to the final apocenter of the GSE galaxy before it was entirely disrupted in the \citet{Naidu+2021} simulations. This prediction has been recently confirmed by the H3 Survey \citep{Han+2022}. Another key feature is that stars beyond about 20\,kpc (the ``outer halo") are very sparsely represented in the local halo relative to their true distribution. 

The lower panel of Figure \ref{fig:r_lz_hist} shows the relative amount of stellar mass as a function of $L_z$ in the outer halo beyond 20\,kpc. A large prograde component from Sagittarius is present in the whole halo which is missing from the local halo. Local halo stars are much more radial than the entire halo; i.e., very small values of $L_z$ are over-represented. This is because, analogous to the Sagittarius dwarf galaxy, stars with higher angular momenta from other dwarfs are less likely to pass near the Sun. 

\begin{figure}[t!]
\centering
\includegraphics[width=1\linewidth]{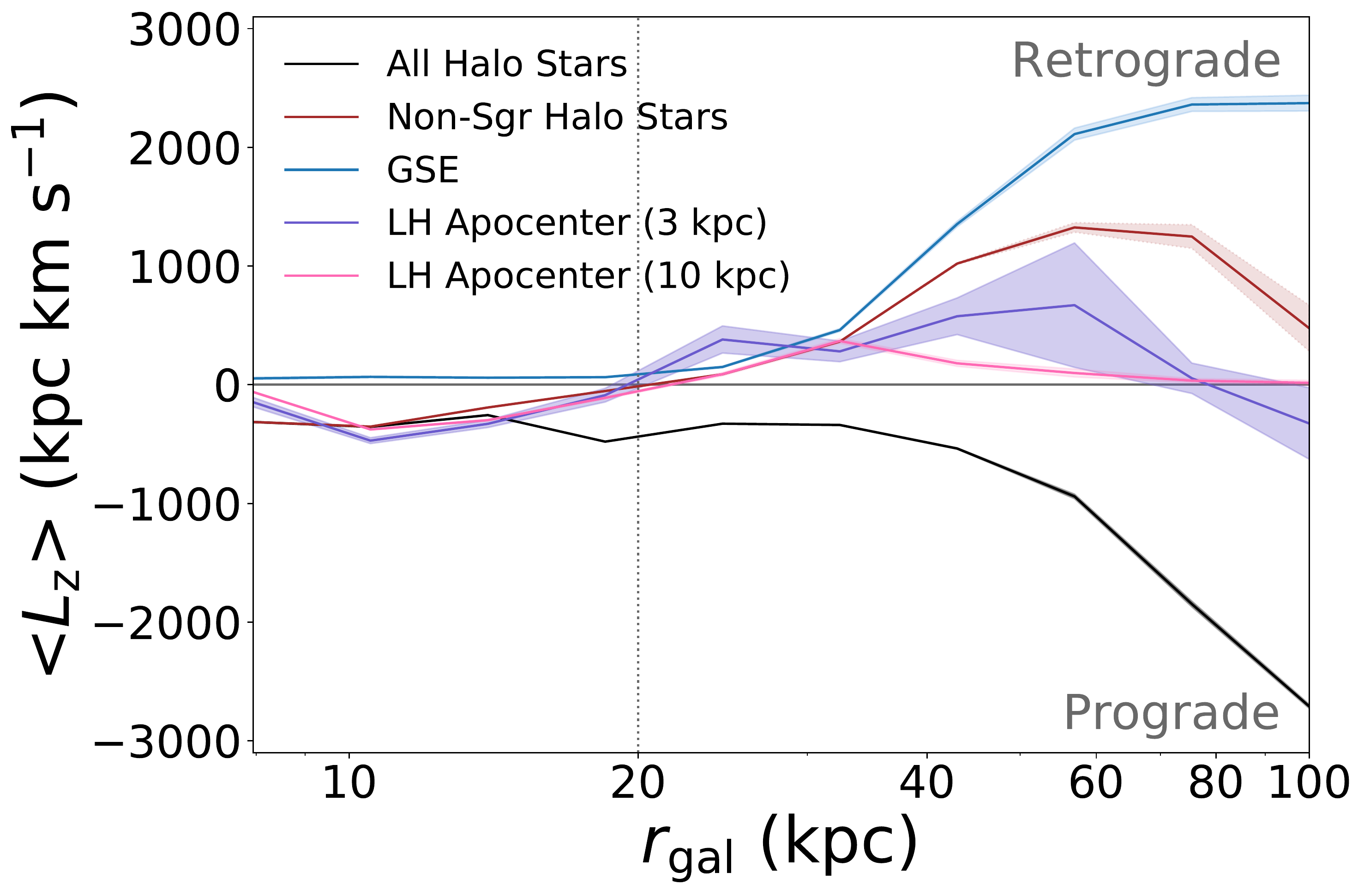}
\caption{Comparison of the average $L_z$ of all halo stars (in black), halo stars excluding Sagittarius (brown), GSE stars (blue), and local halo apocenters (for a solar neighborhood of 3\,kpc in purple). One standard deviation is shaded above and below each line. The gray dotted line marks the beginning of the outer halo (20\,kpc). The angular momentum distribution of the outer halo extrapolated from the local halo is significantly more radial than the actual halo.}
\label{fig:avg_lz_r}
\end{figure}

We examine the prograde/retrograde motion of the halo in more depth by considering $L_z$ as a function of 3D galactocentric distance, $r_{\rm{gal}}$ (Figure \ref{fig:avg_lz_r}). The whole halo is prograde on average out to 100\,kpc, and the magnitude of $L_z$ increases with radius. In the more distant halo, this is due to the contribution from Sagittarius. When we exclude Sagittarius, the net motion in the distant halo is significantly retrograde owing to GSE.

Interestingly, the local halo samples selected from the composite model predict a retrograde outer halo, the magnitude of which is consistent with \citet{Carollo+2007,Carollo+2010}. However, the actual non-Sagittarius halo in the model at these distances is \textit{even more} retrograde. This trend is because the outer-most wraps of GSE at $r_{\rm{gal}}\gtrsim50$ kpc predicted by the \citet{Naidu+2021} simulations, and motivated by the ``Arjuna" substructure associated with GSE \citep{Naidu+2020}, are extremely retrograde. 

\subsection{Comparison to Other Milky Way-like Stellar Halo Simulations}
\label{sec:MW_context}

\begin{figure*}[ht]
\centering
\includegraphics[width=1.0\textwidth]{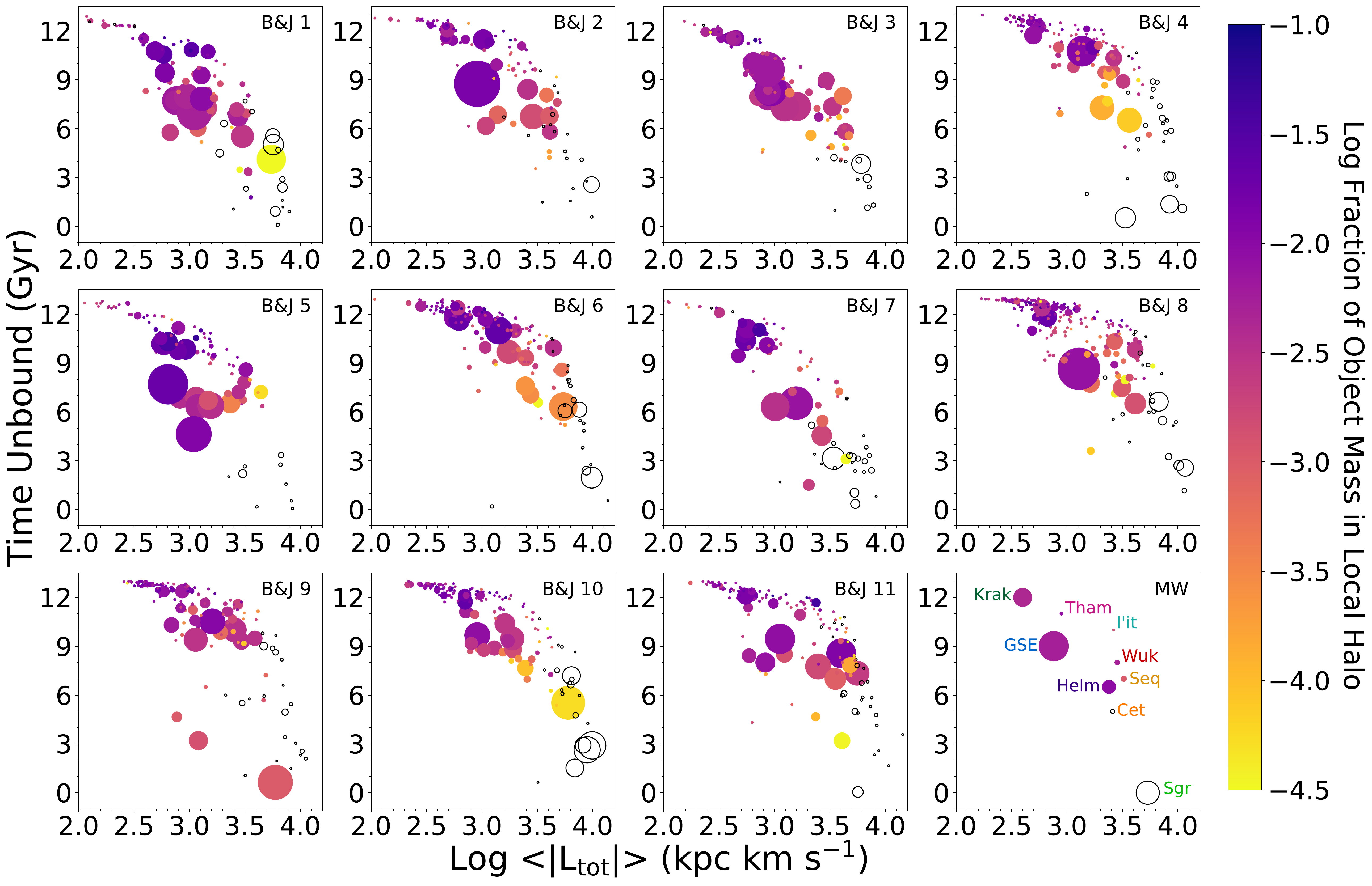}
\caption{Twelve panels showing properties of disrupted dwarfs from each of the eleven \citetalias{Bullock+Johnston2005} Milky Way-like stellar halos that are produced exclusively from dwarf debris, and our composite stellar halo. For each accreted dwarf galaxy, the ``time unbound", or lookback time to when the galaxy ceased to be a gravitationally bound object, is plotted against the logarithm of average angular momentum. Each point, representing a distinct disrupted dwarf galaxy, is colored by the fraction of the object's mass in the 3\,kpc local halo, with black, empty circles having no local halo stars. Circle sizes are proportional to the total object stellar mass to the power 3/4. In the lower right panel, each of the nine halo objects are labeled, with colors consistent to those in Figures \ref{fig:mw_halo_xy}, \ref{fig:mw_halo_xz} and \ref{fig:frac_halo_sn}. Notice that objects with high angular momentum and more recent times unbound --- particularly those with total angular momenta $\gtrsim 4000\,\mathrm{kpc}\,\mathrm{km}\,\mathrm{s}^{-1}$ --- generally have low or no solar neighborhood representation. Additionally, objects at very low angular momenta, $\lesssim 300\,\mathrm{kpc}\,\mathrm{km}\,\mathrm{s}^{-1}$ tend to have lower representation than those with moderate angular momenta.}
\label{fig:3kpcTubLtot}
\end{figure*}

Our composite halo is comprised of simulations representing the nine most massive accreted galaxies which comprise the vast majority of the stellar halo by mass. However, we note that there are several lower mass objects that we do not include in this analysis, which owing to their small halo fractions present only minor perturbations to our findings here. Further, there are likely several disrupted dwarfs that are yet to be discovered \citep[e.g.,][]{Fattahi+2020}. To explore the full possible parameter space, we compare the nine dwarfs within our composite halo to the $\approx1,000$ found within the eleven Milky Way-like stellar halos from \citetalias{Bullock+Johnston2005}.

In Figure \ref{fig:3kpcTubLtot}, we place our composite Milky Way stellar halo in context amongst the eleven \citetalias{Bullock+Johnston2005} Milky Way-like stellar halos. Each panel corresponds to a single stellar halo, within which every accreted dwarf galaxy is plotted as a separate circle, with more massive dwarfs being larger. Points are colored as per the fraction of the dwarf's mass that is represented in the local halo of radius 3\,kpc; dwarfs with no stars within the local halo are shown as empty black circles. 

The accreted dwarf galaxies in our composite halo, when compared to the halos from \citetalias{Bullock+Johnston2005} fall within a similar band of increasing average ${L}_{\mathrm{tot}}$ with decreasing time unbound (the time at which each galaxy ceased to be a gravitationally bound object). Across this band, we see a gradient of representation wherein objects in the bottom right (i.e., high angular momenta, recently accreted objects) are the least likely to be present in the local halo. This pattern is consistent with Sagittarius and Cetus in our composite halo. 

These trends excitingly imply that there is a large population of recently accreted disrupted dwarfs (empty circles in Figure \ref{fig:3kpcTubLtot}) with high angular momenta orbiting almost exclusively at large distances which are yet to be discovered. These galaxies are predominantly low-mass systems, relatively unmixed, and present as coherent streams at $\gtrsim50$ kpc that promise to provide exquisite constraints on the mass distribution (e.g., \citealt{Bonaca+2014,Sanderson17,Vasiliev+2021}) and dynamic disequilibrium in the Galaxy (e.g., due to the Large Magellanic Cloud; \citealt{Garavito-Camargo2019,Conroy+21,Petersen21,Erkal2021,Lilleengen+22}).

To a much lesser extent, dwarfs with very low angular momenta which were accreted very early in the history of the Galaxy ($\gtrsim12-13$\Gyr ago) are also poorly represented in the local halo. For the few accreted dwarfs in this regime which are entirely unrepresented, their debris are concentrated around the galactic center, and therefore beyond the solar neighborhood. However, interestingly a local halo census manages to capture the majority of these most ancient galaxies ingested by the Milky Way.

\section{Conclusions} \label{sec:conclusion}

The overwhelming majority of halo studies rely on solar neighborhood samples. We create a composite Milky Way stellar halo from simulations of the nine most massive disrupted dwarf galaxies to contrast the local halo to the whole halo. Further, we use the eleven stellar halo models built for Milky Way-mass galaxies from \citetalias{Bullock+Johnston2005} to place our composite model in context. Our findings are as follows:

\begin{enumerate}
    \item The local halo does not accurately represent the composition of the stellar halo; some dwarfs are excluded entirely from the sample, while others may be greatly over- or under-represented. [Fig. \ref{fig:frac_halo_sn}, Sec. \ref{sec:local_halo_results}]
    
    \item Extrapolating the properties of the outer halo via orbital integration of local halo stars does not yield an accurate reflection of the outer halo. For example, the strong retrograde rotation of the outer halo (excluding Sagittarius) is underestimated. This is because the most distant halo stars as well as those with the highest angular momentum do not pass through the solar neighborhood. [Fig. \ref{fig:r_lz_hist} and \ref{fig:avg_lz_r}, Sec. \ref{sec:local_halo_results}]
    
    \item Comparing with the \citetalias{Bullock+Johnston2005} simulations, we find that the chief class of disrupted dwarf galaxies entirely missing from the local halo is comprised of recently accreted systems with high angular momentum (e.g., Cetus). These systems are under-represented in our current census of the halo. [Fig. \ref{fig:3kpcTubLtot}, Sec. \ref{sec:MW_context}]
\end{enumerate}

In light of the biases discussed in this work, we urge caution when interpreting local halo samples. The composite model presented in this work may already be used as a realistic approximation of the stellar halo to e.g., model survey selection functions. We envision future work will produce self-consistent simulations including boutique models for all dwarfs.

Our findings motivate whole halo samples that both reach to the edge of the Galaxy as well as into the Galactic center -- efforts in this spirit include the 2MASS M-giant sample from \citealt{Majewski+2003}, the Pan-STARRS RR Lyrae samples from \citealt{Sesar+2017,Cohen+2017}, and the H3 Survey \citep{Conroy+2019}. Upcoming surveys like SDSS-V \citep{SDSSV}, DESI \citep{Prieto20}, 4MOST \citep{4MOSTlowres}, and WEAVE \citep{Dalton12} promise to reveal the most ancient as well as most recent entrants into our Galaxy which are currently missing from our census of disrupted dwarf galaxies.

\software{
    \package{{Astropy} v4.2.1}
    \citep{Astropy-Collaboration+2013,Astropy-Collaboration+2018},
    \package{{gala}} \citep{gala,Price-Whelan+2018},
    \package{{jupyter}} \citep{jupyter},
    \package{{matplotlib}} \citep{matplotlib},
    \package{{numpy}} \citep{numpy}
    }

\begin{acknowledgements}

We thank Nelson Caldwell for feedback on an early version of this project. We thank Eugene Vasiliev, Vasily Belokurov, and Denis Erkal for their simulation of Sagittarius; Helmer Koppelman, Amina Helmi, Davide Massari, Sebastian Roelenga, and Ulrich Bastian for their simulation of the Helmi Streams; Khyati Malhan, Zhen Yuan, Rodrigo Ibata, Anke Arentsen, Michele Bellazzini, and Nicolas Martin for their simulation of Wukong/LMS-1; Jiang Chang, Zhen Yuan, Xiang-Xiang Xue, Iulia Simion, Xi Kang, Ting Li, Jing-Kun Zhao, and Gang Zhao for their simulation of Cetus; and James Bullock, Kathryn Johnston, Andreea Font, Brant Robertson, and Lars Hernquist for their eleven Milky Way-like stellar halo simulations.

Support for this work was provided by NASA through the NASA Hubble Fellowship grant HST-HF2-51515.001-A awarded by the Space Telescope Science Institute, which is operated by the Association of Universities for Research in Astronomy, Incorporated, under NASA contract NAS5-26555.

\end{acknowledgements}

\bibliography{my_bib, my_bib2}
\bibliographystyle{aasjournal}

\end{document}